\documentclass[aps,prd,twocolumn,nofootinbib]{revtex4}
\usepackage{adjustbox}
\usepackage{epstopdf}
\usepackage{graphicx,color}
\usepackage{array}
\newcolumntype{Z}{>{\centering\arraybackslash}m{\dimexpr(\columnwidth-10\tabcolsep-2.4pt)/5\relax}}
\usepackage[colorlinks=true, linkcolor=blue, citecolor=blue, urlcolor=blue]{hyperref}

\begin{document}

\title{Renormalization and Factorization Scale-Invariant Predictions for the $e^+e^-\to J/\psi+\eta_c$ process at $B$ factories }

\author{Si-Qi Gong$^1$}

\author{Ting Sun$^1$}
\email[email:]{tingsuny@gzmu.edu.cn}

\author{Sheng-Quan Wang$^1$}
\email[email:]{sqwang@alu.cqu.edu.cn}

\author{Jian-Ming Shen$^{2}$}
\email[email:]{shenjm@hnu.edu.cn}

\author{Xu-Dong Huang$^{3}$}
\email[email:]{huangxd@cqnu.edu.cn}

\author{Xing-Gang Wu$^4$}
\email[email:]{wuxg@cqu.edu.cn}

\author{Leonardo Di Giustino$^{5,6}$}
\email[email:]{leonardo.digiustino@uninsubria.it}

\address{$^1$Department of Physics, Guizhou Minzu University, Guiyang 550025, P.R. China}
\address{$^2$School of Physics and Electronics, Hunan University, Changsha 410082, P.R. China}
\address{$^3$College of Physics and Electronic Engineering, Chongqing Normal University, Chongqing 401331, P.R. China}
\address{$^4$Department of Physics, Chongqing University, Chongqing 401331, P.R. China}
\address{$^5$Department of Science and High Technology, University of Insubria, via Valleggio 11, I-22100, Como, Italy}
\address{$^6$INFN, Sezione di Milano-Bicocca, 20126 Milano, Italy}

\date{\today}

\begin{abstract}
We perform a NNLO perturbative QCD analysis for exclusive double charmonium production $e^+e^-\to J/\psi+\eta_c$ within the NRQCD factorization formalism using the Principle of Maximum Conformality (PMC). It is well known that this charmonium process suffers poor pQCD convergence, as well as large renormalization scale uncertainty. The PMC is used to remove conventional renormalization scale uncertainty. Based on two fundamental physical processes of charmonium, $J/\psi\rightarrow e^+e^-$ and $\eta_c\rightarrow\gamma\gamma$, and using the NNLO perturbative results together with experimental measurements, we extract the factorization-scale-dependent long-distance matrix elements for $J/\psi$ and $\eta_c$. Combining these two matrix elements with the factorization-scale-dependent short-distance coefficient, we obtain predictions that are independent of the factorization scale for the $e^+e^-\to J/\psi+\eta_c$ process. Our predictions agree with the measurement from the BaBar collaborations within errors. Remarkably, we provide a new way to eliminate conventional renormalization and factorization scale ambiguities for the $e^+e^-\to J/\psi+\eta_c$ process. We further demonstrate that the slow convergence of the pQCD series is an intrinsic property of this double-charmonium process.
\end{abstract}

\maketitle

\section{Introduction}
\label{sec:1}
The exclusive production of double charmonium in electron-positron annihilation, exemplified by the notable process $e^+e^-\to J/\psi+\eta_c$, has long served as a critical testing ground for the interplay between perturbative and non-perturbative QCD. Experimentally, this process was first observed by the Belle collaboration in 2002~\cite{Belle:2002tfa}, followed by more refined measurements from both the Belle and the BaBar collaborations~\cite{Belle:2004abn,BaBar:2005nic}\footnote{Recently, the Belle collaboration has measured the cross section of $e^+e^-\to J/\psi+\eta_c$ at the $\sqrt{s}=10.52$ GeV with an integrated luminosity of $955~fb^{-1}$; however, no evident signal of the double charmonium state is found~\cite{Belle:2023gln}.}. The measured cross sections, however, exhibited a significant discrepancy with the leading-order (LO) QCD predictions~\cite{Braaten:2002fi,Liu:2002wq,Hagiwara:2003cw} based on the nonrelativistic QCD (NRQCD) factorization formalism~\cite{Bodwin:1994jh}, which were found to be an order of magnitude smaller than the experimental measurements. This so-called ``double charmonium puzzle" has stimulated extensive theoretical efforts over the past two decades~\cite{Brodsky:2003hv,Cheung:2003xw,Ma:2004qf,Bondar:2004sv,Braguta:2005kr,Zhang:2005cha,Bodwin:2006dm,Ebert:2006xq,He:2007te,Choi:2007ze,Gong:2007db,Bodwin:2007ga,Guo:2008cf,Mengesha:2011pu,Dong:2012xx,Li:2013otv,Bodwin:2014dqa,Sun:2018rgx,Feng:2019zmt,Sun:2021tma,Zeng:2021hwt,Huang:2022dfw,Li:2025mng,Chen:2025qgy}.

A major breakthrough came with the next-to-leading-order (NLO) QCD corrections~\cite{Zhang:2005cha,Gong:2007db}, which were found to be large and positive, substantially reducing the gap between theory and experiment. Subsequent studies incorporated relativistic corrections at ${\cal O}(v^2)$ and ${\cal O}(\alpha_sv^2)$~\cite{Braaten:2002fi,He:2007te,Dong:2012xx,Bodwin:2007ga,Li:2013otv}, further improving the agreement.

The first complete next-to-next-to-leading-order (NNLO) QCD calculation for $e^+e^-\to J/\psi+\eta_c$ was carried out by Feng \textit{et al.}~\cite{Feng:2019zmt} using numerical master integrals and the auxiliary mass flow method, confirming the validity of NRQCD factorization at NNLO for the first time. Their results showed that the NNLO corrections are sizable, and the renormalization scale dependence is reduced to a certain degree. Shortly after, Huang \textit{et al.}~\cite{Huang:2022dfw} independently reproduced the NNLO corrections, also including bottom-quark loop contributions. More recently, Li \textit{et al.}~\cite{Li:2025mng} derived the asymptotic expansion of the short-distance coefficients (SDCs) up to $r^{15}$, where $r=16m_c^2/s$, allowing for efficient phenomenological applications over a wide range of center-of-mass energies. Their results demonstrated excellent agreement with the full calculation for $r<0.8$, and enabled the conversion between on-shell and $\overline{\rm{MS}}$ mass schemes. Concurrently, Chen \textit{et al.}~\cite{Chen:2025qgy} performed a high-precision analytic calculation of the two-loop amplitudes using differential equations, identifying the origin of leading logarithms and extending the predictions to the $\Upsilon+\eta_b$ system.

Despite these advances, the precision of theoretical predictions remains unsatisfactory, especially owing to the uncertainties arising from both the renormalization scale $\mu_r$ and the factorization scale $\mu_{\Lambda}$. This motivates the need for a more detailed study of the renormalization and factorization scale-setting problem, which is, in fact, an important improvement for pQCD predictions.

In conventional pQCD treatments, the renormalization scale $\mu_r$ for the $e^+e^-\to J/\psi+\eta_c$ process is set to $Q$ and $Q$ is often taken as $2m_c$, $\sqrt{s}/2$, or $\sqrt{s}$. Theoretical uncertainties are subsequently estimated by varying this scale $Q$ within an arbitrary range, commonly specified as $\mu_r\in [Q/2,2Q]$. This conventional procedure violates the fundamental principle of renormalization group invariance (RGI)~\cite{Stueckelberg:1952hma,Gell-Mann:1954yli,Bogolyubov:1956gh,Callan:1970yg,Symanzik:1970rt,Peterman:1978tb} and introduces ambiguities related to the renormalization scheme and scale, thereby inducing substantial uncertainties in pQCD predictions. The conventional scale uncertainty for the $e^+e^-\to J/\psi+\eta_c$ process is still large even for NNLO calculations. More specifically, the renormalization scale errors are $\Delta\sigma_{\rm{LO}}=^{+46\%}_{-30\%}$, $\Delta\sigma_{\rm{NLO}}=^{+45\%}_{-28\%}$, and $\Delta\sigma_{\rm{NNLO}}=^{+52\%}_{-29\%}$ for the production cross sections at LO, NLO, and NNLO, respectively.

The Principle of Maximum Conformality (PMC)~\cite{Brodsky:2011ta,Brodsky:2012rj,Brodsky:2011ig,Mojaza:2012mf,Brodsky:2013vpa} provides a rigorous, RG-compliant framework to eliminate renormalization scale and scheme ambiguities. The PMC method extends the well-known Brodsky-Lepage-Mackenzie (BLM) method~\cite{Brodsky:1982gc} to all orders and reduces to the Gell-Mann-Low method~\cite{Gell-Mann:1954yli} in the Abelian limit~\cite{Brodsky:1997jk}. QCD predictions using PMC satisfy the principle of renormalization group invariance (RGI)~\cite{Brodsky:2012ms,Wu:2014iba,Wu:2018cmb,Wu:2019mky}. Distinct from conventional arbitrary scale prescriptions, the PMC scales are unambiguously determined by absorbing all $\beta$-terms governed non-conformal contributions into the running coupling. In this paper, we present a comprehensive analysis of the $e^+e^-\to J/\psi+\eta_c$ process at NNLO using the PMC method.

\section{PMC scale setting for the $e^+e^-\to J/\psi+\eta_c$ process}
\label{sec:2}
In the NRQCD framework, the total production cross section of the $e^+e^-\to J/\psi+\eta_c$ process factorizes into perturbatively calculable short-distance coefficients (SDCs) and nonperturbative long-distance matrix elements (LDMEs) encoding quarkonium bound-state dynamics, which are conventionally expressed in terms of the radial wave functions at the origin:
\begin{eqnarray}
\sigma &=&\frac{12\alpha^2M_{J/\psi}M_{\eta_c}|\mathbf{P}_1|^3|R_{J/\psi}(0)|^2|R_{\eta_c}(0)|^2}{\pi\,s^{\frac{3}{2}}\,m_c^4}\,|f|^2.
\label{EQ:sigma1}
\end{eqnarray}
where $\alpha$ is the QED coupling constant, $\sqrt{s}$ is the center-of-mass energy, $M_{J/\psi(\eta_c)}$ is the mass of the $J/\psi$ ($\eta_c$) meson, $m_c$ denotes the mass of the charm quark, and $\left|\mathbf{P}_1\right|$ is determined by
\begin{eqnarray}
\left|\mathbf{P}_1 \right| = \frac{\lambda^{\frac{1}{2}}(s, M_{J/\psi}^2, M_{\eta_c}^2)}{2\sqrt{s}},
\label{EQ:P}
\end{eqnarray}
with $\lambda(x, y, z) =x^2 + y^2 + z^2 - 2(xy + xz + yz)$. The short-distance coefficient $f$ is expanded perturbatively in $\alpha_s$, and at $\mu_r=\sqrt{s}/2$ it can be written as:
\begin{eqnarray}
f=f_0+\hat{f_1}\,a_s(\mu_r)+\big(\hat{f_2}+(\gamma_{J/\psi}+\gamma_{\eta_c})\ln\frac{\mu_\Lambda^2}{m_c^2}\big)\,a_s^2(\mu_r),
\end{eqnarray}
where $a_s(\mu_r)=\alpha_s(\mu_r)/(4\pi)$ is the QCD coupling constant. The occurrence of $\ln\frac{\mu_\Lambda^2}{m_c^2}$ is required by the NRQCD factorization and the anomalous dimensions $\gamma_{J/\psi}$ and $\gamma_{\eta_c}$ are given by
\begin{eqnarray}
\gamma_{J/\psi}&=&-\pi^2\left(\frac{C_AC_F}{4}+\frac{C^2_F}{6}\right), \\
\gamma_{\eta_c}&=&-\pi^2\left(\frac{C_AC_F}{4}+\frac{C^2_F}{2}\right),
\end{eqnarray}
the color factors $C_A=3$ and $C_F=\frac{4}{3}$. The coefficients $f_0$, $f_1(\hat{f_1}=f_1|_{\mu_r=\sqrt{s}/2})$, and $f_2(\hat{f_2}=f_2|_{\mu_r=\sqrt{s}/2})$ correspond to the LO, NLO, and NNLO pQCD corrections, respectively, and analytic expressions in the $\overline{\rm{MS}}$ scheme can be obtained from Refs.~\cite{Li:2025mng,Chen:2025qgy}.

The parameters $|R_{J/\psi}(0)|$ and $|R_{\eta_c}(0)|$ are the radial wave functions at the origin for $J/\psi$ and $\eta_c$, respectively. The factorization-scale-dependent radial wave functions is non-perturbative yet universal, which can be determined in a model-independent way from experimental observables and perturbatively calculable quantities, especially, based on two fundamental physical processes of charmonium, $J/\psi\rightarrow e^+e^-$ and $\eta_c\rightarrow\gamma\gamma$, i.e.,
\begin{eqnarray}
\Gamma_{J/\psi\to e^+e^-}&=&\frac{4\, e_c^2\, \alpha^2}{M_{J/\psi}^2}\,|R_{J/\psi}(0)|^2\,\mathcal{C}_{J/\psi}(\mu_r,\mu_{\Lambda}),\label{EQ:C1}\\
\Gamma_{\eta_c\to \gamma \gamma}&=&\frac{3\, e_c^4\, M_{\eta_c}^3\, \alpha^2}{8\, m_c^5}\, |R_{\eta_c}(0)|^2\,\mathcal{C}_{\eta_c}(\mu_r,\mu_{\Lambda}),
\label{EQ:C}
\end{eqnarray}
where $e_c$ denotes the $c$-quark electric charge, and $\mathcal{C}_{J/\psi}(\mu_r,\mu_{\Lambda})$ and $\mathcal{C}_{\eta_c}(\mu_r,\mu_{\Lambda})$ are the short-distance coefficients for $J/\psi$ and $\eta_c$, respectively. Using the NNLO perturbative results~\cite{Beneke:1997jm,Czarnecki:2001zc,Feng:2015uha}, together with experimental measurements, we extract the factorization-scale-dependent radial wave functions $|R_{J/\psi}(0)|^2$ and $|R_{\eta_c}(0)|^2$:
\begin{eqnarray}
|R_{J/\psi}(0)|^2&=&\frac{M_{J/\psi}^2\Gamma_{J/\psi\to e^+e^-}}{4\, e_c^2 \alpha^2\,\mathcal{C}_{J/\psi}(\mu_r,\mu_{\Lambda})}, \label{EQ:rjp}\\
|R_{\eta_c}(0)|^2&=&\frac{8\, m_c^5\,\Gamma_{\eta_c\to\gamma\gamma}}{3\, e_c^4\, M_{\eta_c}^3\, \alpha^2\,\mathcal{C}_{\eta_c}(\mu_r,\mu_{\Lambda})}.
\label{EQ:ret}
\end{eqnarray}
Substituting the extracted wave functions $|R_{J/\psi}(0)|^2$ and $|R_{\eta_c}(0)|^2$ into Eq.(\ref{EQ:sigma1}), the production cross section can be reformulated as
\begin{eqnarray}
\sigma &=&\frac{8\,m_cM_{J/\psi}^3|\mathbf{P}_1|^3\,\Gamma_{J/\psi\to e^+e^-}\,\Gamma_{\eta_c\to\gamma\gamma}}{\pi\,e_c^6M_{\eta_c}^2s^{\frac{3}{2}}\alpha^2}\,R(\mu_r).
\label{EQ:sigma11}
\end{eqnarray}

The radial wave functions $|R_{J/\psi}(0)|^2$ and $|R_{\eta_c}(0)|^2$ determined from Eqs.(\ref{EQ:rjp}) and (\ref{EQ:ret}) evolves with the factorization scale according to its own renormalization group equation (RGE). By inverting the corresponding $\{\beta_i\}$-terms in the perturbative series directly associated with the RGE of the radial wave functions, we can unambiguously extract the physical, factorization-scale-independent component of the series. The leading-order anomalous dimension $\gamma_{J/\psi}$ of the relevant NRQCD vector current is identical for the $e^+e^-\to J/\psi+\eta_c$ and $J/\psi\rightarrow e^+e^-$ processes. Likewise, the anomalous dimension $\gamma_{\eta_c}$ is identical for the $e^+e^-\to J/\psi+\eta_c$ and $\eta_c\rightarrow\gamma\gamma$ processes. The factorization-scale-dependent term $\ln\frac{\mu_\Lambda^2}{m_c^2}$ disappears in Eq.(\ref{EQ:sigma11}), and thus eliminates a dominant source of theoretical uncertainty present in conventional treatments based on potential-model wave functions or arising from the tacit use of a unified LDME regardless of the factorization scale.

The renormalization-scale dependent factor $R(\mu_r)$ takes the form
\begin{eqnarray}
R(\mu_r)&=&c_{1,0}\,a_s^2(\mu_r)+\big(c_{2,0}+c_{2,1}\,n_f\big)\,a_s^3(\mu_r) \nonumber\\
&&+\big(c_{3,0}+c_{3,1}\,n_f+c_{3,2}\,n_f^2\big)\,a_s^4(\mu_r).
\label{EQ:Rmr}
\end{eqnarray}
where $n_f$ denotes the number of active flavors. If we calculate the $e^+e^-\to J/\psi+\eta_c$ process directly in the $\overline{\rm MS}$ scheme, we shall obtain a small PMC scale~\cite{Huang:2022dfw}, which is close to the critical scale $\Lambda_{\rm QCD}$. A proper choice of renormalization scheme is helpful to avoid the small scale problem. For the $e^+e^-\to J/\psi+\eta_c$ process, we transform pQCD series from the $\overline{\rm MS}$ scheme to the momentum space subtraction scheme (MOM scheme)~\cite{Celmaster:1979km,Celmaster:1979dm,Celmaster:1979xr,Celmaster:1980ji}. Unlike the $\overline{\rm MS}$ scheme, the MOM schemes are more physical and carry information about the quark-gluon vertex at a specific momentum configuration. Use of the MOM scheme avoids the complexity arising from the simultaneous evolution of the $\overline{\rm MS}$ mass and the strong coupling constant~\cite{Yan:2024oyb}. One may expect a higher level of convergence for pQCD series of the physical quantities using the physical MOM scheme. After applying the PMC, all the scheme-dependent non-conformal $\{\beta_i\}$-terms have been removed, and the resultant conformal series becomes scheme-independent. The scheme independence can also be ensured by the ``commensurate scale relations"~\cite{Brodsky:1994eh, Lu:1992nt}.

The coupling constant between the $\overline{\rm{MS}}$ scheme and the MOM scheme is established within the framework of perturbation expansion.
\begin{eqnarray}
a_s(\mu_r) = \sum_{i=1}^{\infty} d_i^{\rm{m}} \left(a_s^{\rm{m}}(\mu_r)\right)^i,
\label{EQ:asMOM}
\end{eqnarray}
where $a_s^{\rm{m}}(\mu_r)=\alpha_s^{\rm{m}}(\mu_r)/(4\pi)$ is the QCD coupling constant in the MOM scheme, and the coefficients $d_1^{\rm{m}}$, $d_2^{\rm{m}}$ and $d_3^{\rm{m}}$ are given by~\cite{Gracey:2014pba,Bednyakov:2020cdf}:
\begin{eqnarray}
d^{\rm{m}}_1&=&1, \\
d^{\rm{m}}_2&=&-26.4925+3.4168\,n_f, \\
d^{\rm{m}}_3&=&443.2412-159.9938\,n_f+15.6617\,n_f^2.
\end{eqnarray}
We can transform the $e^+e^-\to J/\psi+\eta_c$ process from the $\overline{\rm{MS}}$ scheme to the MOM scheme using Eq.(\ref{EQ:asMOM}), and it can be written as
\begin{eqnarray}
R(\mu_r)&=&c^{\rm{m}}_{1,0}\,\big(a_{s}^{\rm{m}}(\mu_{r})\big)^2+ \big(c^{\rm{m}}_{2,0} +c^{\rm{m}}_{2,1}n_{f}\big)\,\big(a_{s}^{\rm{m}}(\mu_{r})\big)^3\nonumber\\
&&+\big(c^{\rm{m}}_{3,0}+c^{\rm{m}}_{3,1}n_{f}+c^{\rm{m}}_{3,2}n_{f}^{2}\big)\,\big(a_{s}^{\rm{m}}(\mu_{r})\big)^4.
\label{EQ:RmrMOM}
\end{eqnarray}
The $n_f$ series in Eq.(\ref{EQ:RmrMOM}) is transformed into the $\{\beta_i\}$ series by using the degeneracy relation~\cite{Mojaza:2012mf},
\begin{eqnarray}
R(\mu_r)&=&r^{\rm{m}}_{1,0}\,\big(a_{s}^{\rm{m}}(\mu_{r})\big)^2 + \big(r^{\rm{m}}_{2,0} + 2\beta_{0}r^{\rm{m}}_{2,1}\big)\,\big(a_{s}^{\rm{m}}(\mu_{r})\big)^3 \nonumber\\
&&+\big(r^{\rm{m}}_{3,0} +2\beta_{1}r^{\rm{m}}_{2,1}+3\beta_{0}r^{\rm{m}}_{3,1}+ 3\beta_{0}^{2}r^{\rm{m}}_{3,2}\big) \big(a_{s}^{\rm{m}}(\mu_{r})\big)^4. \nonumber \\
\label{EQ:RmrMOMbe}
\end{eqnarray}
The coefficients $r_{i,0}^{\rm{m}}$ with $i=(1,2,3)$ are scale-invariant conformal contributions, the coefficients $r_{i,j}^{\rm{m}}$ with $1\leq j<i\leq 3$ are non-conformal contributions that are absorbed into the running coupling by shifting the renormalization scale. After using the PMC, we obtain
\begin{eqnarray}
R(\mu_r)&=&r^{\rm{m}}_{1,0}\,\big(a_{s}^{\rm{m}}(Q_\star)\big)^2 + r^{\rm{m}}_{2,0}\,\big(a_{s}^{\rm{m}}(Q_\star)\big)^3 \nonumber\\
&&\quad + r^{\rm{m}}_{3,0}\,\big(a_{s}^{\rm{m}}(Q_\star)\big)^4.
\label{EQ:RmrMOMpmc}
\end{eqnarray}
The PMC scale $Q_\star$ is given by:
\begin{eqnarray}
\ln \frac{Q_\star^2}{(\sqrt{s}/2)^2} = T_0 + T_1 \, a_s^{\text{m}}(\sqrt{s}/2),
\label{EQ:pmcscale}
\end{eqnarray}
where
\begin{eqnarray}
T_0 &=& -\frac{\hat{r}_{2,1}^{\rm{m}}}{\hat{r}_{1,0}^{\rm{m}}}, \\
T_1 &=& \frac{3(\hat{r}_{2,0}^{\rm{m}} \hat{r}_{2,1}^{\rm{m}} - \hat{r}_{1,0}^{\rm{m}} \hat{r}_{3,1}^{\rm{m}})}{2(\hat{r}_{1,0}^{\rm{m}})^2} + \frac{3((\hat{r}_{2,1}^{\rm{m}})^2 - \hat{r}_{1,0}^{\rm{m}} \hat{r}_{3,2}^{\rm{m}})}{2(\hat{r}_{1,0}^{\rm{m}})^2} \beta_0, \nonumber\\
\label{EQ:T}
\end{eqnarray}
where $\hat{r}_{i,j}^{\rm{m}} = r_{i,j}^{\rm{m}}|_{\mu_r=\sqrt{s}/2}$. The conformal coefficients $r^{\rm{m}}_{1,0}$, $r^{\rm{m}}_{2,0}$, and $r^{\rm{m}}_{3,0}$ can be written as
\begin{eqnarray}
r^{\rm{m}}_{1,0}&=&c^{\rm{m}}_{1,0}, \\
r^{\rm{m}}_{2,0}&=&c^{\rm{m}}_{2,0}+\frac{33}{2}\,c^{\rm{m}}_{2,1}, \\
r^{\rm{m}}_{3,0}&=&c^{\rm{m}}_{3,0}+\frac{1}{4}\big(-642\,c^{\rm{m}}_{2,1}+33\,(2\,c^{\rm{m}}_{3,1}+33\,c^{\rm{m}}_{3,2})\big).
\end{eqnarray}
The conformal coefficients are independent of the renormalization scale $\mu_r$.

\section{Numerical results and discussions}
\label{sec:3}
For our numerical calculations, input parameters including the masses and the decay widths of the $J/\psi$ and $\eta_c$ mesons are adopted from the Particle Data Group (PDG)~\cite{ParticleDataGroup:2024cfk}: $M_{J/\psi}=3.09690$ GeV, $M_{\eta_c}=2.98409$ GeV, $\Gamma_{J/\psi\to e^+e^-}=5.53\pm0.10$ keV and $\Gamma_{\eta_c\to\gamma\gamma}=6.4\pm0.5$ keV. The three-loop QCD coupling constant $\alpha_s$ is adopted and its corresponding asymptotic scale $\Lambda_{\rm{QCD}}^{\overline{\rm{MS}}}$ is determined using the world average value $\alpha_s(M_Z)=0.1180$~\cite{ParticleDataGroup:2024cfk}. The asymptotic scale in the MOM scheme is derived via the relation $\Lambda_{\rm{QCD}}^{\rm{m}}=\Lambda_{\rm{QCD}}^{\overline{\rm{MS}}} \exp[-d^{\rm{m}}_2/(2\beta_0)]$~\cite{Celmaster:1979km}. The QED coupling constant is fixed at $\alpha=1/130.9$, and the charm quark mass is taken as $m_c=1.5$ GeV.

\begin{table}[htb]
\centering
\begin{tabular}{|c|c|c|c|c|c|}
\hline
~~$\mu_r$~~    & ~~$\rm{LO}$~~ & ~~$\rm{NLO}$~~ & ~~$\rm{NNLO}$~~ & ~~$\rm{total}$~~ \\
\hline
~~$2\,\rm{GeV}$~~& ~~5.33~~  & ~~7.68~~ & ~~16.48~~ & ~~29.49~~  \\
~~$3\,\rm{GeV}$~~& ~~3.19~~  & ~~4.67~~ & ~~8.25~~ & ~~16.11~~  \\
~~$\sqrt{s}/2$~~ & ~~1.92~~  & ~~2.93~~ & ~~4.57~~ & ~~9.42~~  \\
~~$\sqrt{s}$~~   & ~~1.26~~  & ~~2.03~~ &~~ 3.02~~ & ~~6.30~~  \\
\hline
\end{tabular}
\caption{The LO, NLO, and NNLO QCD correction terms (in fb) for the $e^+e^-\to J/\psi+\eta_c$ process at $\sqrt{s}=10.58$ GeV using the conventional scale setting.}
\label{tb.sec}
\end{table}

In Table \ref{tb.sec} we present the LO, NLO, and NNLO QCD correction terms for the $e^+e^-\to J/\psi+\eta_c$ process at $\sqrt{s}=10.58$ GeV using the conventional scale setting. With the same input parameters, our results in the $\overline{\rm{MS}}$ scheme are in agreement with those presented in Ref.~\cite{Li:2025mng}. For the case of fixing the renormalization scale $\mu_r=\sqrt{s}/2$ and varying the scale $\mu_r\in[2m_c,\sqrt{s}]$ in the $\overline{\rm{MS}}$ scheme, the scale $\mu_r$ uncertainties are $\Delta\sigma_{\rm{LO}}=^{+46\%}_{-30\%}$, $\Delta\sigma_{\rm{NLO}}=^{+45\%}_{-28\%}$, and $\Delta\sigma_{\rm{NNLO}}=^{+52\%}_{-29\%}$ for the total production cross sections at LO, NLO, and NNLO, respectively. After transforming the $e^+e^-\to J/\psi+\eta_c$ process from the $\overline{\rm{MS}}$ scheme to the physical MOM scheme, Table \ref{tb.sec} shows that the scale $\mu_r$ uncertainties are $\Delta\sigma_{\rm{LO}}=^{+66\%}_{-35\%}$, $\Delta\sigma_{\rm{NLO}}=^{+62\%}_{-32\%}$, and $\Delta\sigma_{\rm{NNLO}}=^{+71\%}_{-33\%}$ for the total production cross sections at LO, NLO, and NNLO, respectively. In the MOM scheme the scale uncertainty is not reduced by the inclusion of higher-order QCD calculations, and the predictions are still plagued by significant scale uncertainties at the present NNLO accuracy.

\begin{figure}[htb]
\centering
\includegraphics[width=0.45\textwidth]{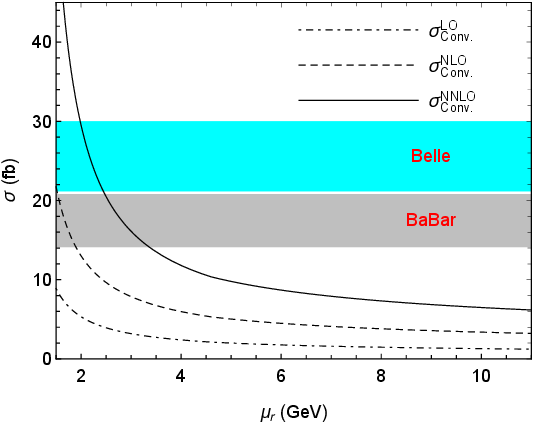}
\caption{Production cross sections for $e^+e^-\to J/\psi+\eta_c$ versus the renormalization scale $\mu_r$, calculated in the physical MOM scheme with conventional (Conv.) scale setting. The dot-dashed, dashed, and solid curves show the central values of the LO, NLO, and NNLO predictions, respectively. The bands show the measurements reported by the Belle~\cite{Belle:2004abn} and BaBar~\cite{BaBar:2005nic} collaborations.}
\label{Fig:Conv}
\end{figure}

More explicitly, in Fig.(\ref{Fig:Conv}) we present the production cross sections of $e^+e^-\to J/\psi+\eta_c$ versus the renormalization scale $\mu_r$ in the physical MOM scheme using conventional scale setting. The production cross sections of $e^+e^-\to J/\psi+\eta_c$ at the LO and NLO depend heavily on the renormalization scale $\mu_r$, and the scale uncertainty is not reduced by the inclusion of the NNLO QCD correction. The sensitivity to the renormalization scale is more pronounced at lower values of the scale and gradually diminishes with increasing $\mu_r$. We can see from Table \ref{tb.sec} and Fig.(\ref{Fig:Conv}) that the quality of the convergence of the pQCD series depends heavily on the choice of renormalization scale, and one cannot decide the intrinsic convergence of the pQCD series, and a poor convergence of the pQCD series may be caused by the improper choice of the renormalization scale $\mu_r$.

In accordance with the PMC principle, different types of RGEs require separate treatment in order to obtain the correct effective values for the corresponding scale-dependent running parameters. It is noted that the $n_f$ term in the light-by-light contribution is free of ultraviolet divergences and is thus unrelated to the renormalization of the strong coupling. This term should be treated as a part of the conformal contribution in the determination of $\alpha_s$ when the PMC method is applied. In addition, $n_f$-dependent terms arise from other scale-dependent quantities, such as long-distance matrix element, this term should also be treated as conformal contribution in the determination of $\alpha_s$. After properly identifying the $n_f$ term, and applying the PMC, we obtain the PMC scale
\begin{eqnarray}
Q_\star=7.06~\rm {GeV}
\end{eqnarray}
for the $e^+e^-\to J/\psi+\eta_c$ process. The PMC scale $Q_\star$ is independent of the renormalization scale $\mu_r$. We note that the obtained PMC scale is between $\sqrt{s}/2$ and $\sqrt{s}$ with $\sqrt{s}=10.58$ GeV. After applying the PMC, the resulting scale-independent conformal coefficients $r_{2,0}$ and $r_{3,0}$ are significantly different from the conventional coefficients. In the case of conventional scale setting, the expansion expression of the production cross section is as follows,
\begin{eqnarray}
\sigma&=&29.85\big(\alpha_{s}^{\rm{m}}(\mu_r)\big)^2+178.89_{-44.90}^{+54.87}\big(\alpha_{s}^{\rm{m}}(\mu_r)\big)^3 \nonumber\\
&&+1098.64_{-375.06}^{+595.93}\big(\alpha_{s}^{\rm{m}}(\mu_r)\big)^4,
\end{eqnarray}
where the uncertainty of perturbative coefficients is caused by varying the scale $\mu_r\in[2m_c,\sqrt{s}]$. The expansion expressions using the PMC is
\begin{eqnarray}
\sigma&=&29.85\big(\alpha_{s}^{\rm{m}}(Q_\star)\big)^2+315.79\big(\alpha_{s}^{\rm{m}}(Q_\star)\big)^3 \nonumber\\
&&+2418.56\big(\alpha_{s}^{\rm{m}}(Q_\star)\big)^4.
\end{eqnarray}
At NLO and NNLO, the scale-independent conformal coefficients using the PMC are larger than the conventional coefficients in the range of $\mu_r\in[2m_c,\sqrt{s}]$.

\begin{table}[htb]
\centering
\begin{tabular}{|c|c|c|c|c|c|}
\hline
 & ~~$\rm{LO}$~~ & ~~$\rm{NLO}$~~ & ~~$\rm{NNLO}$~~ & ~~$\rm{total}$~~ \\
\hline
~~$\rm{PMC}$~~ & ~~1.59~~ & ~~3.89~~ & ~~6.89~~ & ~~12.38~~  \\
\hline
\end{tabular}
\caption{The LO, NLO, and NNLO QCD correction terms (in fb) for the $e^+e^-\to J/\psi+\eta_c$ process at $\sqrt{s}=10.58$ GeV using the PMC scale setting. The PMC predictions are independent of the renormalization scale $\mu_r$.}
\label{tb.secPMC}
\end{table}

\begin{figure}[htb]
\centering
\includegraphics[width=0.45\textwidth]{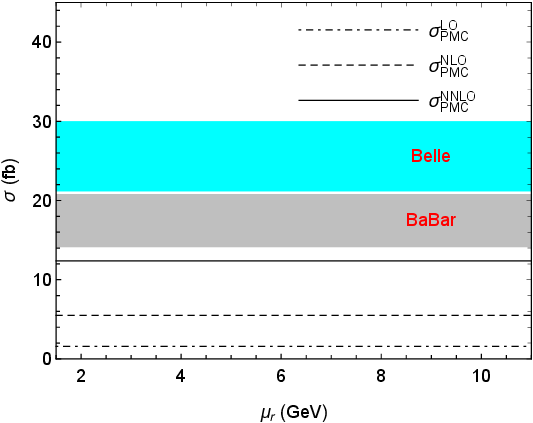}
\caption{Production cross sections for $e^+e^-\to J/\psi+\eta_c$ versus the renormalization scale $\mu_r$ with PMC scale setting. The dot-dashed, dashed, and solid curves show the central values of the LO, NLO, and NNLO predictions, respectively. The bands show the measurements reported by the Belle~\cite{Belle:2004abn} and BaBar~\cite{BaBar:2005nic} collaborations.}
\label{Fig:PMCscale}
\end{figure}

In Table \ref{tb.secPMC} we present the LO, NLO, and NNLO QCD correction terms for the $e^+e^-\to J/\psi+\eta_c$ process at $\sqrt{s}=10.58$ GeV using the PMC scale setting. For the current $e^+e^-\to J/\psi+\eta_c$ process up to NNLO, the PMC scale $Q_{\star}$ is independent of the choice of scale $\mu_r$, and so are the conformal coefficients $r^{\rm{m}}_{i,0}$, thus the resulting pQCD predictions eliminate the renormalization scale ambiguities. The PMC predictions for the NLO and NNLO QCD correction terms are larger than the conventional results at $\mu_r=\sqrt{s}/2$ and $\sqrt{s}$, which indicates the slow convergence of the pQCD series for the $e^+e^-\to J/\psi+\eta_c$ process. More explicitly, in Fig.(\ref{Fig:PMCscale}) we present the production cross sections of $e^+e^-\to J/\psi+\eta_c$ versus the renormalization scale $\mu_r$ using PMC scale setting. The production cross sections eliminate the scale uncertainty for each order as well as for the total cross section. In general, after absorbing all non-conformal $\beta$ terms into the QCD running coupling so that the divergent renormalon terms disappear from the series, the PMC is expected to make the perturbative expansion more convergent. This advantage is not yet fully reflected in the current $e^+e^-\to J/\psi+\eta_c$ process; this suggests that the slow convergence at NNLO is an inherent characteristic for the $e^+e^-\to J/\psi+\eta_c$ process, indicating the importance of uncalculated NNNLO QCD corrections\footnote{As a brief estimate based on the Pad$\acute{e}$ approximation approach (PAA)~\cite{Basdevant:1972fe, Samuel:1992qg, Samuel:1995jc}, the NNNLO term is found to be $10.42$ fb for [N/M]=[0/2] and $12.19$ fb for [N/M]=[1/1], the resulting production cross sections are consistent with the Belle and BaBar measurements.}. It is expected that the effect of PMC in improving convergence will become more pronounced with the inclusion of higher-order QCD calculations.

For the $e^+e^-\to J/\psi+\eta_c$ process, in addition to the uncertainties from the renormalization and factorization scales, another important source of uncertainty originates the $c$-quark mass $m_c$. Moreover, the value of the QCD coupling constant $\alpha_s(M_Z)$ also introduces additional uncertainties. Although our present analysis removes the uncertainty associated with the factorization scale $\mu_\Lambda$, conventional results still suffer from substantial renormalization scale $\mu_r$ uncertainty. After applying the PMC, the conventional renormalization scale $\mu_r$ ambiguity is eliminated. The production cross sections for $e^+e^-\to J/\psi+\eta_c$ are
\begin{eqnarray}
\sigma|_{\mathrm{Conv.}}&=& 9.42\pm 0.17\pm 0.74^{+1.58+0.53+6.69}_{-1.44-0.49-3.11} \ \mathrm{fb}, \\
\sigma|_{\mathrm{PMC}}&=& 12.38\pm 0.22\pm 0.97^{+2.08+0.69}_{-1.89-0.65} \ \mathrm{fb},
\end{eqnarray}
where the first error arises from the experimental uncertainty of $\Gamma_{J/\psi\to e^+e^-}$, the second is from the experimental uncertainty of $\Gamma_{\eta_c\to\gamma\gamma}$, the third is from the $c$-quark mass $\Delta m_c=0.1$ GeV, and the fourth is from the coupling constant $\Delta\alpha_s(M_Z)=0.0009$. In the case of conventional scale setting, the fifth error is caused by varying the scale $\mu_r\in[2m_c,\sqrt{s}]$, which gives the dominant contribution to the uncertainty of the production cross section. Finally, the total uncertainties are $\Delta\sigma|_{\rm{Conv.}}=^{+74\%}_{-38\%}$ for the conventional prediction and $\Delta\sigma|_{\rm{PMC}}=^{+19\%}_{-18\%}$ for the PMC prediction. Since the PMC eliminates the renormalization scale uncertainty, the precision of our PMC predictions is greatly improved.

We note that the recent PDG value for the $\eta_c\to\gamma\gamma$ decay is $\Gamma_{\eta_c\to\gamma\gamma}=6.4\pm0.5$ keV~\cite{ParticleDataGroup:2024cfk}, which differs significantly from the 2025 PDG result of $\Gamma_{\eta_c\to\gamma\gamma}=5.1\pm0.4$ keV~\cite{ParticleDataGroup:old}. Hence, the experimental determination of this decay width remains to be further refined. The BESIII Collaboration has recently reported a decay width of $\Gamma_{\eta_c\to\gamma\gamma}=7.48\pm1.48\pm0.30$ keV~\cite{BESIII:2026pff} for this process. Taking this representative value of $\Gamma_{\eta_c\to\gamma\gamma}=7.48\pm1.48\pm0.30$ keV as the input, we obtain the production cross sections for $e^+e^-\to J/\psi+\eta_c$ as follows\footnote{We note that the decay width was measured to be $\Gamma_{\eta_c\to\gamma\gamma}=11.30\pm0.56\pm0.66\pm1.14$ keV by BESIII in 2025~\cite{BESIII:2024rex}. Using this measured value as the input, we derive a production cross section of $\sigma|_{\rm{PMC}}=21.85\pm 0.40\pm 2.77^{+3.67+1.21}_{-3.34-1.15}$ fb for the $e^+e^-\to J/\psi+\eta_c$ process, in agreement with the Belle and BaBar measurements.}:
\begin{eqnarray}
\sigma|_{\mathrm{Conv.}}&=& 11.01\pm 0.20\pm 2.22^{+1.85+0.61+7.82}_{-1.68-0.58-3.63} \ \mathrm{fb}, \\
\sigma|_{\mathrm{PMC}}&=& 14.47\pm 0.26\pm 2.92^{+2.43+0.80}_{-2.21-0.76} \ \mathrm{fb}.
\end{eqnarray}

\begin{figure}[htb]
\centering
\includegraphics[width=0.45\textwidth]{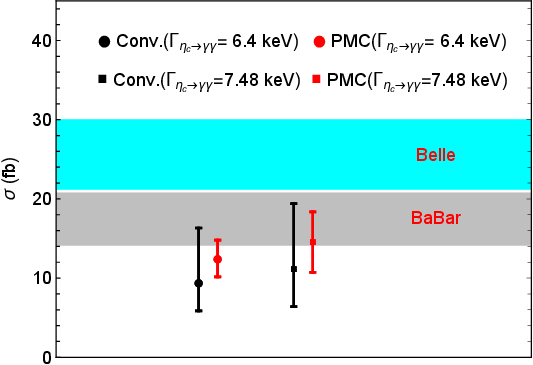}
\caption{Production cross sections for the $e^+e^-\to J/\psi+\eta_c$ process. Two representative values of the $\eta_c\to\gamma\gamma$ decay width, $\Gamma_{\eta_c\to\gamma\gamma}=6.4$ keV and $\Gamma_{\eta_c\to\gamma\gamma}=7.48$ keV, are taken as input parameters. Measurements from the Belle~\cite{Belle:2004abn} and BaBar~\cite{BaBar:2005nic} collaborations are shown for comparison.}
\label{Fig:totcp}
\end{figure}

We present the production cross sections for the $e^+e^-\to J/\psi+\eta_c$ process in Fig.(\ref{Fig:totcp}), where the errors denote the combined uncertainties from $\Delta\Gamma_{J/\psi\to e^+e^-}$, $\Delta\Gamma_{\eta_c\to\gamma\gamma}$, the $c$-quark mass $\Delta m_c=0.1$ GeV, the coupling constant $\Delta\alpha_s(M_Z)=0.0009$ and the renormalization scale uncertainty $\mu_r\in[2m_c,\sqrt{s}]$. Our predictions agree
with the measurement from the BaBar~\cite{BaBar:2005nic} collaborations within errors.

\section{Summary}
\label{sec:4}
The conventional practice is to set the renormalization scale to a characteristic scale $Q$, often taken as $2m_c$, $\sqrt{s}/2$, or $\sqrt{s}$; theoretical uncertainties are subsequently estimated by varying the scale $Q$ within an arbitrary range for the pQCD prediction of the $e^+e^-\to J/\psi+\eta_c$ process. The conventional predictions are plagued by significant renormalization scale uncertainties at NNLO. The quality of the convergence of the pQCD series depends heavily on the choice of renormalization scale, and one cannot decide the intrinsic convergence of the pQCD series for the $e^+e^-\to J/\psi+\eta_c$ process.

In this paper, we provide a detailed PMC analysis for the $e^+e^-\to J/\psi+\eta_c$ process, demonstrating that using the RGEs to determine the correct effective values of scale-dependent running parameter provides a mechanism to eliminate the standard factorization-scale dependence. All the non-conformal $\beta$ terms are absorbed in the determination of $\alpha_s$ and only the conformal contributions are retained in the pQCD series. The PMC predictions eliminate the renormalization scale uncertainty. Thus, we obtain a consistent higher-order pQCD prediction free of both renormalization-and factorization-scale ambiguities for the $e^+e^-\to J/\psi+\eta_c$ process. Furthermore, we demonstrate that the slow convergence at NNLO is an inherent characteristic for the $e^+e^-\to J/\psi+\eta_c$ process, indicating the importance of uncalculated NNNLO QCD corrections. The scale-independent predictions are
\begin{eqnarray}
\sigma|_{\rm{PMC}}&=&12.38^{+2.41}_{-2.23} \,\,\rm{fb}, \nonumber \\
\sigma|_{\rm{PMC}}&=&14.47^{+3.89}_{-3.75} \,\,\rm{fb} \nonumber
\end{eqnarray}
for $\Gamma_{\eta_c\to\gamma\gamma}=6.4$ keV and $\Gamma_{\eta_c\to\gamma\gamma}=7.48$ keV, respectively, and are in good agreement with the measurement from the BaBar~\cite{BaBar:2005nic} collaborations. This analysis shows the importance of the application of a reliable renormalization scale setting method in perturbative QCD and our predictions are helpful for detailed studies of properties of the heavy quarkonium processes.

\hspace{1cm}

{\bf Acknowledgements}: This work was supported in part by the Natural Science Foundation of China under Grant No.12265011, No.12575080 and No.12547101; by the Project of Guizhou Provincial Department under Grant No.CXTD[2025]030, No.2025YJSKYJJ170, and No.GZMUBS6; by the Hunan Provincial Natural Science Foundation with Grant No.2024JJ3004, YueLuShan Center for Industrial Innovation (2024YCII0118); by the Natural Science Foundation of Sichuan Province under Grant No.2024NSFSC1367.

\end{document}